\documentstyle[twocolumn,floats,aps]{revtex}
\begin{document}
\draft

\twocolumn[\hsize\textwidth%
\columnwidth\hsize\csname@twocolumnfalse\endcsname

\title{\bf The Coulomb Luttinger liquid}

\author{D. W. Wang$^{(1)}$, A. J. Millis$^{(2)}$, and S. Das Sarma$^{(1)}$}

\address{
(1)Department of Physics, University of Maryland,
College Park, Maryland 20742-4111\\
(2)Center for Materials Theory and Department of Physics and
Astronomy, Rutgers University, New Brunswick, New Jersey 08554\\
}

\date{\today}
\maketitle
\pagenumbering{arabic}


\begin{abstract}
Accurate expressions, valid in experimentally relevant regimes,
are presented for the
effect of long-ranged Coulomb interaction on the low energy properties 
(momentum distribution function, density of states, electron
spectral function, and
$4k_F$ correlation function) of one-dimensional electron systems.
The importance of plasmon dispersion (as opposed to exponent) effects
in the spectral function is demonstrated.
\end{abstract}

\pacs{PACS numbers:71.10.-w; 71.10.Pm; 74.20.Mn; 73.20.Mf; 73.20.Dx.}
\vskip 1pc]
\narrowtext
The low energy behavior of one-dimensional (1D) electron systems is 
known not to be consistent with Fermi liquid theory~\cite{Luther72}. 
However, the theoretically well established~\cite{Haldane83} and 
extensively studied~\cite{Review} Luttinger liquid model of one-dimensional 
physics \cite{footnote} is,
strictly speaking, not applicable to electronically 
conducting one-dimensional systems such as quantum wires (QWR)
\cite{QWRref}, carbon nanotubes~\cite{nanotube}, organic conductors
\cite{organic} and doped chain or ladder compounds~\cite{chain}, because
the electrons in these compounds interact via the Coulomb force,
which is long-ranged, whereas the standard Luttinger model assumes 
short-ranged interaction. The long range of the Coulomb interaction leads
to a scale dependence of the Luttinger exponents and velocities
\cite{schulz}, which have been studied by several authors~\cite{schulz,leading}
on the assumption that it is well approximated by its leading ($\ln^{1/2}$)
divergence. As we show, this approximation is not accurate in 
any physically relevant regime. One exception is a very interesting recent
renormalization group treatment~\cite{RG} which found an effective 
exponent very similar to ours but did not discuss the implications 
for physical quantities. Some numerical results have also
appeared~\cite{lattice}, but a general understanding of the experimental 
implications of the Coulomb interaction is lacking.

In the present paper we use direct analytical and numerical evaluation of
the relevant bosonization expressions to determine
the momentum distribution function, tunneling
density of states, and spectral function for 1D electron systems 
interacting via the physically 
relevant Coulomb interaction at zero temperature.
We define an important but previously overlooked energy scale, present 
an accurate expression for the scale dependent exponent, show how the
scale dependent velocity affects the spectral function, 
and qualitatively
discuss the $4k_F$ correlation function.
Our results should apply directly to 1D QWRs~\cite{QWRref} and nanotubes
\cite{nanotube}.

We consider a 1D electron system with a noninteracting
dispersion $\varepsilon _{p}$ which we linearize near the Fermi point,
defining a bare velocity $v_{F}.$ We here assume the only important 
interaction is the Coulomb interaction in the forward scattering channel,
and neglect Umklapp scattering and other interactions. This is a good 
approximation for QWR and nanotube systems. (For the organic and doped spin
chain materials a modification, discussed below Eq. (2), is needed.)
The Hamiltonian is (here we do not write the spin index explicitly),
\begin{eqnarray}
H&=&\sum _{r,p}v_{F}(p-rp_{F})c^\dagger_{r,p}c_{r,p} \nonumber\\
&&+\frac{1}{L}\sum_{r,q}V_{c}(q)
\left(\rho_{r}(q)\rho_{r}(-q)+\rho_{r}(q)\rho_{-r}(-q)\right),
\end{eqnarray}
where $c^\dagger_{r,p}$ is the electron creation operator and $\rho_{r}(q)$
is the density operator describing density fluctuations at
momentum $q$ and branch $r=\pm 1$ for the right(left) movers.
For 1D systems $V_{c}\left(q\right)\rightarrow \ln(1/q)$
as $q\rightarrow 0$, and becomes $1/q$ for $q$ larger
than some scale $q_{0}$ set by the geometry and the wave function size.
A reasonable 
approximate form, which we will use in our subsequent analysis, is
\begin{equation}
V_{c}(q)=\frac{\pi v_{F}V_0}{2}\ln \left[ \frac{q_{0}+q}{q}\right],
\end{equation}
where $V_0$ is a dimensionless measure of the interaction strength and 
$q_0^{-1}$ is the length scale parameter.
$V_0$ and $q_{0}$ are system dependent factors.
For a cylindrical quantum wire of radius $a$, 
$V_0=4e^{2}/\pi\varepsilon_{0}v_{F}$ and $q_0\sim 2.5/a$, where
$e$ is the electron charge, and $\varepsilon _{0}$ is the background
dielectric constant, about 10 for GaAs. 
These values give the correct long wavelength limit 
and are within 10$\%$ of 
the correct $1/q$ coefficient at large momentum.
In carbon nanotubes $V_0$ 
is of the same form as in QWR but $\varepsilon _{0}\sim 1.4$ 
\cite{nanotube} and $q_0\sim 2.97/R$, where $R$ is the radius of the tube.
For organics or doped spin chains, 
additional short-ranged exchange interactions may be important. The usual
arguments \cite{Haldane83} show that these interactions lead, at low energies,
to an additive constant term in $V_c(q)$.

Eq. (1) may be bosonized as usual \cite{Haldane83,Review}; the charge
excitations are plasmons with dispersion
$\omega_q=qv_q$, and velocity
$v_q\equiv v_F\sqrt{1+2V_c(q)/\pi v_F}$ is
\begin{equation}
v_q=v_{F}\sqrt{1+V_0\ln \left[\frac{q_{0}+q}{q}\right]}.
\end{equation}
(Note that we have 
$\lim_{q\rightarrow\infty}\omega_q\sim qv_F+V_0q_0/2\neq qv_F$ for 
Coulomb interaction).
The electron Green function $G_{r}(x,t)\equiv\langle
\psi_r(xt)\psi_r^\dagger(00)\rangle$ is 
\begin{equation}
G_{r}(x,t)=\lim_{\epsilon\rightarrow 0}\frac{e^{irk_Fx}}{2\pi}
\frac{i\exp \left[-\Phi_{r}\left(x,t\right)\right]}
{x-rv_{F}t+i\epsilon},
\end{equation}
where the phase function $\Phi_r(x,t)$ is
\begin{eqnarray}
\Phi_{r}(x,t)&=&\frac{1}{2}\int_{0}^{\infty }\frac{dp}{p}e^{-\epsilon p}
\left(e^{ip(x-rv_{F}t)}-e^{ip(x-rv_pt)}\right) \nonumber\\
&&+2\sinh^2(\theta _{p})\left( 1-\cos
(px)e^{-irpv_pt}\right).
\end{eqnarray}
The exponent parameter $\theta_{p}$ is defined by
\begin{equation}
e^{-2\theta _{q}}=\sqrt{1+V_0\ln\left(\frac{q_{0}+q}{q}\right)}
\sim \sqrt{V_0}\ln^{1/2}\left(\frac{q_s}{q}\right),
\end{equation}
where $q_s\equiv q_0e^{1/V_0}$ and the last approximation is good at 
long wavelengths, $q\ll q_0$.

We now use Eqs. (4)-(6) to study electronic quantities. We begin with 
the momentum distribution function
\begin{equation}
n_r(\delta p)=\frac{-i}{2\pi}\int_{-\infty }^{\infty }dx\frac{e^{-i\delta px}}
{rx-i\epsilon}\exp \left[-\Phi_{r}\left(x,0\right)\right],
\end{equation}
where $\delta p\equiv p-rk_F$. In a noninteracting Fermi gas,
$n_r(\delta p)=\theta(-r\delta p)$. For a short-ranged
LL, the generally accepted result \cite{Voit} is that in the vicinity of
the Fermi momentum, $0.5-n_r(\delta p)\sim \rm{sgn}\it (r\delta p)
\times \{C_{\rm{1}\it}|\delta p|+C_{\rm{2}\it}|\delta p|^{\gamma}\}$ 
with $\gamma$ a LL exponent and  $C_1$ and $C_2$ two
constants. The first term is the
non-critical background coming from high 
energies, while the second (critical) term comes from low energies where LL
physics is important. For the long-ranged interacting model we now consider, 
attention to 
the singularity structure of the noninteracting electron Green function
leads to (let $\delta p>0$ and $r=+1$)
\begin{eqnarray}
n(\delta p)&=&\frac{1}{2}-\frac{1}{\pi }\int_{0}^{\infty }\frac{dx}{x}\sin
\left( \delta px\right) \exp [-\Phi (x,0)] \nonumber\\
&=&\frac{1}{2}+
C_1'\delta p+C_2'\left(\frac{\delta p}{q_0}\right)^{\gamma_q(\delta p)}
+\mbox{higher orders},
\end{eqnarray}
where again the non-singular $C_1'$ term is from the integration over small $x$,
while the singular $C_2'$ term comes from integration over large $x$ and
is a weak function of $\ln^{1/2}(1/\delta p)$. 
The scale dependent exponent $\gamma_q(\delta p)$ is found to be
\begin{eqnarray}
\gamma_q(q)
&\sim&
\frac{1}{2}\left(\frac{1}{3}e^{-2\theta_q}+e^{2\theta_q}-1\right)
\nonumber\\
&\sim&
\frac{\sqrt{V_0}}{6}\ln^{1/2}\left(\frac{q_s}{q}\right)
+\frac{\ln^{-1/2}\left( q_s/q \right)}{2\sqrt{V_0}}-\frac{1}{2}.
\end{eqnarray}
\begin{figure}

 \vbox to 6.5cm {\vss\hbox to 5.cm
 {\hss\
   {\includegraphics{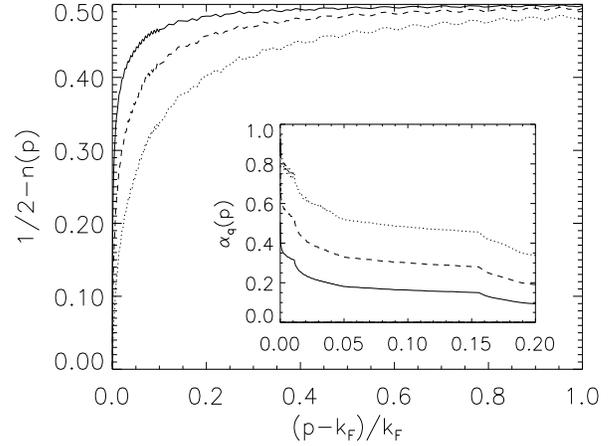}
   }
  \hss}
 }
\caption{
Calculated momentum distribution function, $n(p)$, with respect to momentum
$p-k_F$ for a realistic QWR system of $a=70$ nm. Solid, dashed and dotted lines
are results for three different interaction strengths $V_0=1.21, 2.42, 4.84$ 
respectively, where $V_0=1.21$ is for electron density $0.65\times 10^6$
cm$^{-1}$ and $\varepsilon_0=12.7$ [5]. 
Inset: the effective exponent (around $k_F$) obtained
by taking the logarithmic derivative of the numerical 
$n(p)$ for $|p-k_F|<0.2 k_F$. 
}
\end{figure}

Fig. 1 shows results obtained by numerically evaluating Eq (8) for typical
QWR parameters. An enhanced curvature near the
Fermi momentum is evident. The inset to Fig. 1 shows the logarithmic
derivative $\alpha_q(p)\equiv d\log |n(p)-1/2|/d\log (p)$, which shows that for
small $\delta p$ the behavior may be described in terms of a slowly changing
effective exponent.
We note $\alpha_q(p)$ is always less than 1,
because when the scale dependent exponent $\gamma_q(p)$ of Eq. (9)
is greater than 1, the background term dominates.

We now turn to the tunneling density of states, 
\begin{equation}
N(\omega )=\frac{1}{2\pi}\sum_r\int_{-\infty }^{\infty }dt\;
e^{i\omega t}\left[
G_r(0,t)+G_r(0,-t)\right],
\end{equation}
for $\omega$ measured from the chemical potential.
\begin{figure}

 \vbox to 6.5cm {\vss\hbox to 5.cm
 {\hss\
   {\includegraphics{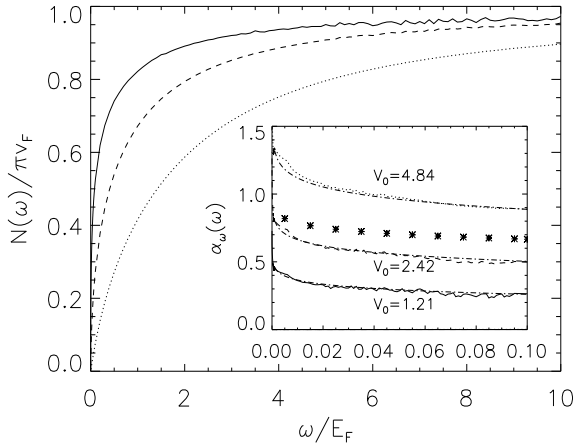}
   }
  \hss}
 }
\caption{
Calculated density of states, $N(\omega)$, with respect to energy $\omega$
for the same system as Fig. 1. Different line styles represent
different interaction strengths as indicated. Inset is the effective
exponent, $\alpha_{\omega}$
obtained by taking the logarithmic derivative of $N(\omega)$. The
numerically calculated curves are well
fitted by the analytical expression (dash-dot lines) of the exponent
from Eq. (13) at the corresponding $V_0$'s and
$\omega_s=20q_sv_F\sqrt{V_0}$. The stars are the
first order term of Eq. (13) only, for $V_0=1.21$, showing
that the widely used leading logarithm approximation leads to factor of
two errors.
}
\end{figure}
We first show that $N(\omega)$ vanishes faster than any power of $\omega$
as $\omega\rightarrow 0.$ We observe that if $V_c(p)\neq 0,$ $G$ vanishes
faster than any power of $t$ as $t\rightarrow \infty$ \cite{schulz}.
Therefore the integral
obtained by taking any number of $\omega$-derivatives of $N(\omega )$ is
absolutely convergent at long times, and may be evaluated straightforwardly
by contour methods even at $\omega =0$ \cite{su}.
We further note that $G_r(0,t)$ has no singularities in the lower
(upper) half plane for $r=+1(-1)$; thus by
deforming the contours appropriately we find that $d^{n}N(\omega )/d\omega
^{n}|_{\omega=0}=0$ for any $n$. This argument does not apply to
$n(p)$ because of the different analytic structure of the $x$-dependence.
Thus the non-critical contributions which obscured the behavior of $n(p)$ do
not occur in $N(\omega)$. By evaluating Eq. (10) we obtain
\begin{equation}
N(\omega)\propto \left(\frac{\omega}{\omega_s}\right)^{\gamma_\omega(\omega)},
\end{equation}
where the scale dependent density of states
exponent $\gamma_\omega(\omega)$ is
\begin{equation}
\gamma_\omega(\omega)\sim
\frac{\sqrt{V_0}}{6}\ln^{1/2}\left(\frac{\omega_s}{\omega}\right)
+\frac{\ln^{-1/2}\left(\omega_s/\omega\right)}{2\sqrt{V_0}}
-\frac{1}{2}.
\end{equation}
the same form as that of Eq. (9) with $q_s$ replaced by a characteristic
energy scale $\omega_s$.
From Eqs. (3) and (6) we expect $\omega_s=A\,q_sv_F\sqrt{V_0}$
with the numerical constant $A$ determined by subleading corrections to the
asymptotic analysis of Eq. (5). $A$ may in principle have a weak scale and
system-parameter dependence, but our numerical results show that for a wide
range of energies ($10^{-3}<\omega/E_F<0.1$) and interactions ($1<V_0<5$)
it is very well approximated by the constant value $A=20$.
Fig. 2 shows the results of a numerical calculation of $N(\omega)$ from
Eq. (10) for three different interaction strengths; the
inset compares the numerically calculated effective exponent,
$\alpha_{\omega}(\omega)\equiv d\log(N(\omega))/d\log(\omega)$, with the
analytical result obtained from Eqs. (11) and (12):
\begin{equation}
\alpha_\omega(\omega)=
\frac{\sqrt{V_0}}{4}\ln^{1/2}\left(\frac{\omega_s}{\omega}\right)
+\frac{\ln^{-1/2}\left(\omega_s/\omega\right)}{4\sqrt{V_0}}
-\frac{1}{2}.
\end{equation}
One sees that the fit is very good (the small differences appearing at
$\omega/E_f\sim 0.01$ arise from noise in the numerical calculation).

The two crucial energy scales defined by $N(\omega)$
are $\omega_s$, and $\omega_\ast$ at which
$\alpha_\omega(\omega_\ast)=1$, corresponding to $\omega_\ast\sim
\omega_se^{-34/V_0}\ll\omega_s$.
In the high energy region, $\omega>\omega_s$, one has essentially non-interacting behavior. For $\omega_\ast<\omega<\omega_s$, one has a LL
with a scale dependent exponent. For $\omega<\omega_\ast$, $\alpha_\omega>1$
and $N(\omega)$ is concave upwards at small $\omega$, suggesting
a "pseudo-gap" in the electronic density of states.
For most real QWR systems, $V_0$ is about 1-5 depending on $\varepsilon_0$
and $v_F$, and thus $\omega_\ast$ is typically many
orders of magnitude 
smaller than $\omega_s$. In our calculation, using QWR parameters
from Ref. \cite{QWRref}, we have $\omega_s\sim 100$ meV and 
$\omega_\ast\sim 10^{-4}$ meV. 
For extremely small $\omega\ll \omega_\ast$, Eq. (13) gives
$\alpha_\omega\sim\frac{\sqrt{V_0}}{4}\ln^{1/2}(\omega_s/\omega)$, 
an approximate form used earlier
in the literature \cite{schulz,leading}. 
However, as seen from the inset of Fig. 2, the leading logarithmic 
divergence is so weak that in all physically
relevant regimes the other two terms in Eq. (12) are needed for quantitative 
accuracy. On the other hand, 
the constant (scale independent) exponent used in Ref. \cite{nanotube}
for nanotubes is also not an adequate approximation for small
energy region ($\omega<0.05 E_F$) either. We therefore propose
Eqs. (9) and (12) as widely applicable fitting formulae for
the effective exponents in the Coulomb Luttinger liquid. 

The scale dependent exponent also appears in the single particle
spectral function, $\rho(q,\omega)=(1/2\pi)[G(q,\omega)+G(-q,-\omega)]$,
however the scale dependent velocity in Eq. (5) is more important.
To introduce our results, we briefly summarize known results for a short-ranged
repulsive interaction in the spinless LL model~\cite{Voit}. At fixed $q$,
one defines three $\omega$-ranges:
(i) $\rho(q,\omega)=0$ for $|\omega|<\omega_q$ (energy-momentum conservation),
and (ii) power-law singularities as $|\omega|\rightarrow\omega_q^+$, and (iii)
an exponential decay at scales larger than the Luttinger cut-off.
For the long-ranged Coulomb interaction, $\rho(q,\omega)=0$ for
$|\omega|<\omega_q$ due to the energy-momentum conservation,
but the behavior in both
regions (ii) and (iii) are strongly modified.
For $|\omega|>\omega_s$ (region (iii)),
$\rho(q,\omega)\sim\exp[-|\omega|/E_c(\omega)]$
with a scale dependent cut-off
\begin{equation}
E_c(\omega)=\frac{q_0v_FV_0}{4}\ln\left(\frac{a\,|\omega|}{v_F}\right),
\end{equation}
because of the slow ($1/q$) decay of the Coulomb interaction
in the large momentum region (Eq. (5)).

Near threshold ($\omega_q<|\omega|\ll\omega_s$) there are two effects: the
scale dependence of the effective Luttinger exponent and the curvature of the
plasmon dispersion, which prevents the different boson modes from adding
coherently. Thus as one decreases $\omega$ towards $\omega_q$ (consider
$\omega>0$ part only) one obtains first
a divergence $\delta\omega^{\gamma_\omega(\delta\omega)-1}$ (here
$\delta\omega\equiv\omega-\omega_q$). This divergence is cut off by curvature
effects at a scale $\omega_c(q)\equiv\mbox{Max}_{p<q}(\omega_p-p\omega_q/q)
\approx (1/4)qv_F\sqrt{V_0}\ln^{-1/2}(q_s/q)$, the difference between the exact
dispersion and a linear approximation. We find that for $q$ larger than
$q_\ast\sim Aq_se^{-75/V_0}$
(at which $\gamma_\omega(\omega_c(q_\ast))=1$)
the curvature effect is more important
in cutting off the divergence, whereas for $q<q_\ast$ the effective exponent is
more important. As $\delta\omega\rightarrow 0^+$ the spectral function
decreases rapidly, ultimately vanishing faster than any power of
$\delta\omega$ due to the increase of the effective exponent. Thus the
generic behavior is a spectral function which increases rapidly as $\omega$
is increased above threshold $\omega_q$, goes through a maximum at
$\omega_{peak}=\omega_q+\Delta\omega$ with $\Delta\omega$ set by the larger
of $\omega_\ast$ and $\omega_c(q)$, and then decreases exponentially
with a scale dependent cut-off $E_c(\omega)$
for $\omega>\omega_{peak}$. The suppressed spectral
weight in the near threshold region is compensated by the slower decay
at high energies, preserving the sum rule $\int\rho(q,\omega)d\omega=1$.
In Fig. 3 we show the results of direct numerical evaluation of the
electron spectral function for the Coulomb Luttinger liquid (solid lines)
and for a short-ranged-interaction (regular) Luttinger liquid with exponent
$\alpha=0.2$, approximately equal to the effective exponent of 
the Coulomb case at $\omega=0.3 E_F$ (dashed lines).
Note that the partition theory techniques used in 
\cite{partition} to simplify the evaluation for the short-ranged case do not
work in the Coulomb case. The shift of the peak away from the threshold 
is evident.

\begin{figure}

 \vbox to 10.5cm {\vss\hbox to 5.cm
 {\hss\
   {\includegraphics{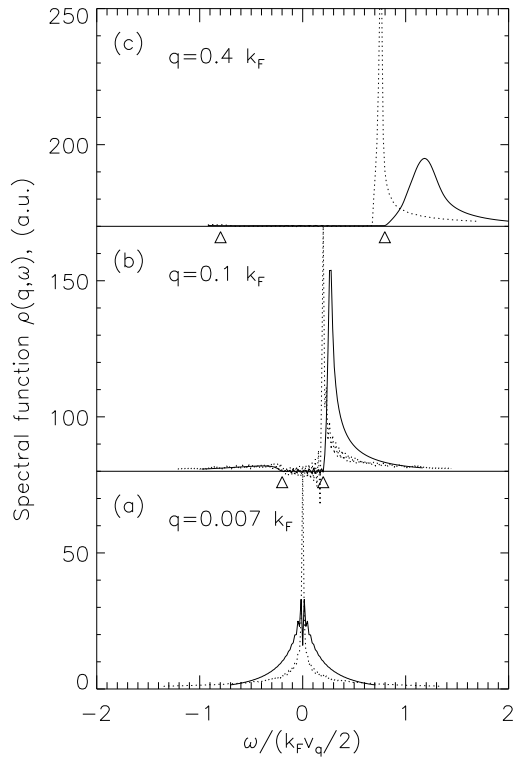}
   }
  \hss}
 }
\caption{
Calculated electron spectral function, $\rho(q,\omega)$, for different
momenta, $q$, as indicated in the figure. Solid lines are from the
Coulomb interacting system (parameters are the same as Fig. 1 with $V_0=1.21$)
while dotted lines are from short-ranged
interacting system with the approximate effective exponent $\alpha=0.2$.
The two triangles
in (b) and (c) indicate the threshold,
$\omega=\pm\omega_q$.
The ripple of the spectral function curves is numerical error.
}
\end{figure}
Finally, we briefly discuss the "Wigner crystal" correlation.
Schultz \cite{schulz} observed that at long enough length scales
the logarithm arising from the
long-ranged Coulomb interaction causes the $4k_F$ component of the
density-density correlation to decay
more slowly than $x^{-4K_\rho}$ and also 
more slowly than the $2k_F$ component, leaving a state best 
interpreted as Wigner crystal.
Using the notations of this paper, we obtain for
the $4k_F$ term in the structure factor,
\begin{eqnarray}
S_{4k_F}(\delta p)\sim\left(\frac{q_s}{\delta p}\right)^
{1-8{V_0}^{-1/2}\ln^{-1/2}(q_s/\delta p)},
\end{eqnarray}
where $\delta p\equiv|p-4k_F|$. Therefore we expect
to see the $4k_F$ divergence when 
$\sqrt{V_0}\ln^{1/2}(q_s/\delta p)>8$ or $\delta p<q_s e^{-64/V_0}$,
or in term of temperature at  
$T<T_{w.x}=\omega_se^{-64/V_0}$, which is sensitive to the
electron density and experimental geometry, but is in general far
too small to be experimentally relevant, and is also much less than
the scale $\omega_\ast$ at which $N(\omega)$ develops a pseudo-gap.

In conclusion, we have presented a systematic theoretical analysis of the
low energy properties of electron systems subject to long-ranged Coulomb 
interactions, including a reliable estimate of the scale dependent 
Luttinger parameter and apparently the first calculation of 
Coulomb effects
on the spectral function, and values for the (unfortunately 
extremely low) scales at which the divergent
behavior associated with the Coulomb interaction becomes manifest. 

This work is supported (DWW and SDS) by US-ONR, US-ARO, and DARPA, and by
NSF-DMR-00081075 (AJM).

\end{document}